\documentclass[aps,prl,preprint]{revtex4}

\begin{document}
\title{Metastabilities in vortex matter}
\author{P Chaddah and S B Roy}
\address{Low Temperature Physics Lab, Centre for Advanced Technology, Indore 452013.}
\begin{abstract}
We extend the classical theory for supercooling across first order phase transitions to the case when both density 
and temperature are control variables. The observable region of metastability then depends on the path followed in 
this space of two variables. Since the density of vortex matter in superconductors can be easily varied over a wide 
range by varying applied field, it is ideal for experimental tests. We found, in our studies on the `peak  effect' in the 
mixed state of superconducting CeRu$_2$, supercooled states whose observable region of metastability depends on the path 
followed in (H,T) space, consistent with our predictions. We also discuss phenomena in hard superconductors that are 
well understood within Bean's critical state model. We conclude that the path dependence of metastablity associated 
with hindered kinetics may be opposite to that predicted for metastability associated with supercooling across a first 
order transition.
\end{abstract}

\maketitle

\section{Introduction}

The mixed state of type II superconductors is seen for applied 
fields H lying between the lower (H$_{C1}$) and upper (H$_{C2}$) 
critical fields. It consists of vortices carrying quantized flux 
which ideally form a two-dimensional hexagonal lattice under 
repulsive forces. Can this lattice of vortices undergo structural 
transitions? Can vortex structures show metastabilities seen in 
usual condensed matter? These questions assume a broader 
significance because the density of vortex matter in 
superconductors can be varied over a wide range ( from 
nearly zero to about 10$^{12}$ per cm$^2$) by varying applied 
magnetic field, and may thus provide better experimental 
tests for metastabilities around phase transitions.

     The period after the discovery of high-T$_C$ superconductors 
(HTSC) has seen many theoretical works proposing vortex-
matter phase transitions\cite{1}. Vortex lattice melting as the field 
(or temperature) is raised towards the H$_{C2}$(T) line now stands 
established as a first order phase transition and experiments 
have established a latent heat, as well as a jump in 
equilibrium magnetisation, satisfying the Clausius-Clapeyron 
relation \cite{2,3,4}.

     Simultaneously, an early theoretical prediction of a first 
order phase transition in the vortex lattice of paramagnetic 
superconductors\cite{5,6,7}, in which the infinitely long vortices get 
segmented into short strings with a sudden enhancement of 
pinning\cite{8}, (and thus critical current density J$_C$ vs H shows a 
peak) has been motivating experimentalists into studying in 
great detail the "peak effect" (PE) in CeRu$_2$. The first 
thermodynamic signature indicating that the onset of the PE is 
a first order transition (FOT), consistent with the theoretical 
prediction of Fulde, Ferrel and Larkin, Ovchinnikov (FFLO), 
came through the observation that the PE appears at a field 
H$^*_a$ on increasing field, but vanishes at a lower field H$^*_d$ on 
decreasing fields\cite{9,10,11}. This hysteresis in the occurrence of the 
PE was taken as the hysteresis expected in a FOT. We have 
attempted to identify other measurable signals of a FOT. The 
vortex matter in CeRu$_2$ has been our paradigm, and the FFLO 
theory has been motivating us, possibly as a red herring. The 
theory is correct and is still used by theorists in understanding 
coexisting superconductivity and (weak) magnetism\cite{12}. Since 
our experiments cannot probe the microscopic nature of the 
phase in the PE region of CeRu$_2$, we shall not discuss the 
relevance of FFLO theory to CeRu$_2$ any further in this talk; 
those interested can see our recent papers\cite{13,14}. 

     In this talk we shall briefly outline the existing wisdom of 
experimental tests for a FOT, and then present our extension 
to the case where one can interchangeably vary two control 
parameters to traverse the FOT line. The need for this 
extension was necessitated by our studies on CeRu$_2$.  We 
shall state new predictions, and discuss experimental 
verification.

\section{Supercooling across first order phase transitions}

 A phase transition is defined as an nth-order transition in the 
Ehrenfest scheme\cite{15} if the nth derivatives of the free energy are 
discontinuous, whereas all lower derivatives are continuous, 
at the transition point. (The derivatives are taken with respect 
to the control variables.) The derivative with respect to 
temperature is entropy and its discontinuity in a FOT implies 
a latent heat, while the derivative with respect to pressure is 
volume which should show a discontinuous change at T$_C$. The 
latent heat and volume change are further related by the 
Clausius-Clapeyron equation.

     The Ehrenfest scheme is ambiguous\cite{15} for some phase 
transitions - one example being the lambda transition in liquid 
helium. Phase transitions are now classified using an order 
parameter S that changes across the phase boundary. The 
change is discontinuous, from $S=0$ to $S=S_0$ for a FOT, but 
continuous for a second-order transition. Two phases can 
coexist at the transition point of a FOT. This is put on a 
formal footing by writing the free energy as a function of the 
order parameter S. When the control variable (say T) 
corresponds to the transition point (T$_C$), then the free energy 
f(S) has two equal minima (at $S=0$ and $S=S_0$) for a  FOT\cite{16}, 
while there is only one minimum for a second order transition. 
One can obviously show that a FOT is accompanied by a 
latent heat and a sudden volume change,  consistent with the 
Ehrenfest scheme.
 
 The existence of two equal minima implies the coexistence of 
two phases at the transition point; slightly away from the 
transition point one still has two minima - one global and 
one local - with slightly unequal values of  the free energy 
f. We show in fig. 1a schematic of  f(S) curves as the 
control variable (T) is varied from above to below the 
transition point (T$_C$). The high temperature phase has higher 
entropy and is `disordered', having an order parameter $S = 0 $. 
while the low temperature phase has a finite (but T--dependent)
order parameter. Since $S=0$ continues to 
correspond to a local minimum in f(S) slightly below T$_C$, one 
can supercool the higher entropy state below the transition 
point\cite{16}. Similarly one can superheat the ordered phase 
above the transition point. One thus sees another 
experimental characteristic of a FOT, viz. the possibility of 
hysteresis in the transition point as one varies a control 
variable. This was the observation\cite{9,10,11} that led to the 
inference that the onset of PE in CeRu$_2$ is a FOT. 

     Concentrating on supercooling, we note from fig. 1 that 
the barrier f$_B$(T) in f(S) separating the metastable state at $S=0$ 
from the stable ordered state reduces continuously as T is 
lowered below T$_C$, and vanishes at the limit of metastablilty 
(T$^*$) of the supercooled state\cite{16}. Supercooling is easily 
observed across the water-ice transition\cite{17}, a FOT familiar to 
all of us, and we believe that the hysteresis in the onset of PE 
in CeRu$_2$ is also a manifestation of the same\cite{18,19}. If the system 
is in the disordered state at T$<$T$_C$, then nucleation of the 
metastable ordered phase occurs, with $f_B(T) >>  kT $, only by 
introducing localised fluctuations of large energy e$_f$. The 
nucleation rate is extremely sensitive to the height of the 
barrier f$_B$, and carefully purified metastable liquids evolve 
suddenly from apparent stability to catastrophic growth of the 
ordered phase\cite{17}. The barrier vanishes below T$^*$, and the 
unstable disordered state now relaxes into the ordered state by 
the spontaneous growth of long-wavelength fluctuations of 
small amplitude, i.e. by spinodal decomposition\cite{17}. (Here we 
shall assume that the ordered stable phase is formed fast 
compared to experimental time scales if   $f_B(T) \leq [e_f+kT] $, and 
the system remains in the metastable state if f$_B$(T) is larger. 
We shall briefly initiate a discussion on kinetics and kinetic 
metastabilities in the last section of this paper).

     Both the water-ice transition, and the onset of PE in 
CeRu$_2$, have been studied extensively with density as a second 
control variable. While the density of water has been varied 
by varying pressure up to 3kbar\cite{17}, the density of vortices is 
varied by varying the applied magnetic field, and the onset of 
PE in CeRu$_2$ has been tracked from 1 Tesla to 4 Tesla\cite{11}, 
corresponding to a four-fold change in vortex density. We can 
now talk of supercooling the disordered $S=0$ phase, at 
differing densities, below the T$_C$(P) corresponding to that 
density. Can one compare the extent of metastability in such 
supercooled states? Secondly, we can cross the transition line 
T$_C$(P) by varying density rather than by varying temperature. 
The f(S) curves are defined once a (T,P) point is defined; the 
$S=0$ state would be metastable just below the FOT line 
irrespective of whether the line is crossed by varying T or by 
varying density. We can thus supercool the disordered phase 
into the region below the FOT line even by varying density. 
Can one compare the metastability in a supercooled state, at a 
point (T,P) below the FOT line, as depending on the path 
followed to reach this (T,P) point? Before pursuing this we 
must emphasize that such questions cropped up in our studies 
on CeRu$_2$ because it is experimentally easy to follow arbitrary 
paths in density (magnetic field) and temperature space in the 
case of vortex matter. The disordered phase here is 
characterised by a larger critical current density J$_C$ 
compared to the ordered phase; supercooling is confirmed 
by measuring the minor hysteresis loops\cite{14,18,19} in contrast to 
the case of supercooled water where one measures 
diffusivity\cite{17}. 

     We have recently argued that while reduction of 
temperature at constant density does not a priori cause 
building up of fluctuations, the very procedure of varying 
density introduces fluctuations\cite{20}. Lowering temperature 
isothermally can keep e$_f$ zero in `carefully purified liquids'. 
Density variation at constant T, however, builds up e$_f$ even in 
such systems. It was noted that free energy curves should be 
plotted for three parameters as f(P,T,S) where P is a generic 
pressure that implies magnetic field in the case of vortex 
matter. Supercooling along various paths in (T,P) space 
involves moving from a f(P$_1$,T$_1$,S) curve to f(P$_2$,T$_2$,S) curve in 
this multidimensional space. These curves have two equal 
minima for (T,P) values lying on the FOT line T$_C$(P), and the 
barrier f$_B$ vanishes for (T,P) values lying on a line T$^*$(P). We 
refer to this T$^*$(P) line as the limit of metastability on 
supercooling.  The first point to note is that while f(P,T,0) is 
weakly dependent on T, it depends strongly on P\cite{20}. The 
second point is that the S-dependence of f(P,T,S), for fixed T, 
is different at different P. This originates from the different 
densities of the ordered and disordered phases, and this was 
also incorporated  by us\cite{20}. Finally, we argued that if density is 
varied then a fraction of the energy change f(P$_1$,T,0) - f(P$_2$,T,0) 
will be randomised into a fluctuation energy e$_f$. This 
last point\cite{20} looks obvious in the case of vortex matter where 
vortices get pinned and unpinned as they move, and the 
energy dissipated in the process is easily measured as the area 
within the M-H loop.  With these physical inputs, we could 
make the following predictions\cite{20}:
\begin{enumerate}
\item  When T$_C$ falls with rising density as in vortex matter 
FOT, or in water-ice transition below 2 kbar, then ($T_C-T^*$) 
will rise with rising density. If  T$_C$ rises with rising density as 
in water-ice above 2 kbar, and in most other solid-liquid 
transitions,  then  ($T_C - T^*$) will fall with rising density. This 
prediction is consistent with known data on water\cite{17}. 
\item  The disordered phase can be supercooled upto the limit 
of metastability T$^*$(P) only if T is lowered in constant P. If the 
T$_C$(P) line is crossed by lowering P at constant T, then 
supercooling will terminate at T$_0$(P) which lies above the 
T$^*$(P) line. If T$_C$ falls with rising density, then 
($T_0(P)-T^*(P)$) rises with rising density.
\item  A supercooled metastable state can be transformed into 
the stable ordered state by density variations through variation 
of pressure or magnetic field. These variations produce a 
fluctuation energy e$_f$ which, when large enough, can cause a 
jump over the free energy barrier f$_B$. In vortex matter e$_f$ is 
related to the area under the M-H loop as the field is varied by 
h. This area, and thus e$_f$, increase monotonically but 
nonlinearly with h. If this field is varied n times with fixed h, 
then e$_f$ will increase linearly with n. With this basic idea, one 
can predict the effect of field variation h on various 
supercooled metastable states\cite{21}. We show, in fig. 2, three 
points in (T,P) space where supercooled states are produced 
by lowering T at constant field. It follows that if h$_0$ is the 
lowest field excursion (with $n=1$) for which the metastable 
state is transformed into a stable state, then h$_0$ will be 
smallest for point 1, and largest for point 3. Further if one 
uses a field variation h$_i$ which is lower than the smallest h$_0$, 
but makes repeated excursions until the stable state is formed 
after n$_0$ such field excursions, then it follows that n$_0$ will be 
smallest for point 1 and largest for point 3.
The predictions made above are qualitative and based on 
general arguments\cite{20,21}; such predictions can be made for 
various possible paths of crossing the T$_C$(P) line. We have 
experimentally confirmed most of the predictions stated above 
with the PE in CeRu$_2$ as our paradigm FOT\cite{22}. 
\end{enumerate}

\section{Hindered kinetics and kinetic metastabilities.}

The experimental confirmation of a FOT involves 
measurement of a volume discontinuity, also of a latent heat, 
and of these two satisfying the Clausius-Clapeyron equation. 
For vortex matter in CeRu$_2$ the discontinuity in vortex volume 
was observed by us\cite{19} but was tedious because we were 
extracting equilibrium magnetisation from hysteretic M-H 
curves\cite{23}.  The latent heat has so far not been measurable, and 
hysteresis was invoked\cite{9,10,11} as a signature of an FOT. We have 
made predictions on path-dependence of metastabilities 
associated with an FOT, and these have also been 
observed.  We must recognise that while we have advanced 
our understanding of metastabilities associated with FOTs, 
metastability can also be kinetic in origin. We wish to now 
address this and pose some questions.
 
     Glasses are known to be metastable, but differ 
significantly from supercooled liquids\cite{24}. The diffusivity of a 
supercooled liquid does not drop suddenly below T$_C$; its 
diffusivity is large enough to permit it to explore configuration 
space on laboratory timescales. The ergodic hypothesis is 
valid, entropy is a valid concept and free energy can be 
defined, permitting the arguments we made in Section 2. A 
glass on the other hand is characterised by low diffusivity and 
hindered kinetics (with a viscosity greater than 10$^{13}$ poise). It 
sits in a local minimum of only the energy landscape and not 
of the free energy, and is non-ergodic\cite{24}. The low 
diffusivity of 
a glass causes metastabilities; the metastabilities are 
associated with hindered kinetics and not with local minima 
in free energy. Hindered kinetics (with kinetic hysteresis) will 
be seen wherever diffusivities are low, examples are critical 
slowing down near a second order phase transition and, closer 
home, M-H hysteresis in hard superconductors where the 
pinning, or 
hindered kinetics, of vortices prevents decay of shielding 
currents (Bean's critical state model). Does a metastability 
induced by hindered kinetics also depend on the path 
followed in (T,P) space? If the metastability is due to 
reduced diffusivity, then na\"ive arguments suggest that the 
metastability will be more persistent when larger motions (of 
particles in configuration space) are involved. And larger 
motions are involved when density is varied, rather than when 
temperature is varied. For the case of vortex matter, a much 
larger rearrangement of vortex structure is involved when we 
reach an (H,T) point by varying field isothermally, than when 
we reach that point by varying temperature at constant field. 
Hysteresis would thus be lower in the field-cooled case, for 
hard superconductors, than in the case of isothermal field 
variation. This is consistent with observations\cite{25}, and with 
predictions of the Bean's critical state model\cite{26}. 
It is also well 
known that Bitter patterns generally show an almost disorder-
free vortex lattice on reducing T in constant H, in striking 
contrast to the case when H is reduced in constant T. We thus 
conclude, with our na‹ve arguments, that the path-dependence of metastability associated with hindered 
kinetics may be opposite to the case of metastability 
associated with a FOT. 

\section{Acknowledgement}
We gratefully acknowledge helpful discussions with Dr. S. M.~Sharma,
 Dr. S. K. Sikka, Dr. Srikanth Sastry, Prof. Deepak Dhar 
and Dr. Sujeet Chaudhary.

\newpage
\begin{center}\bf Figure Captions \end{center}
Fig. 1: We show schematic free energy curves for (a) $T=T^{**}$, 
(b) $T_C<T<T^{**}$, (c) $T=T_C$, (d) $T^*<T<T_C$, and (e) $T=T^*$.

Fig. 2: We show a schematic of the phase diagram with 
supercooled states at 1, 2, and 3 obtained by lowering 
T in constant field (or `pressure' ).

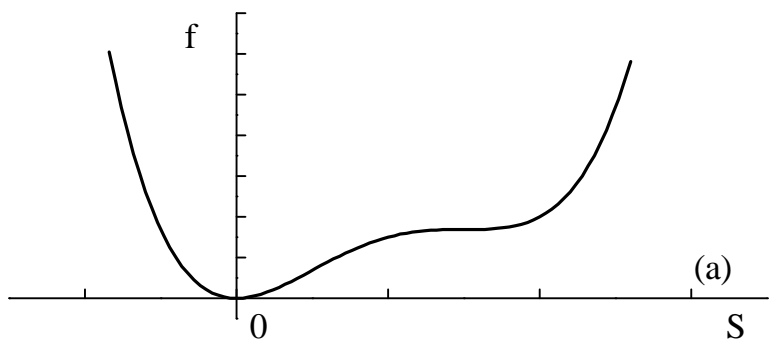
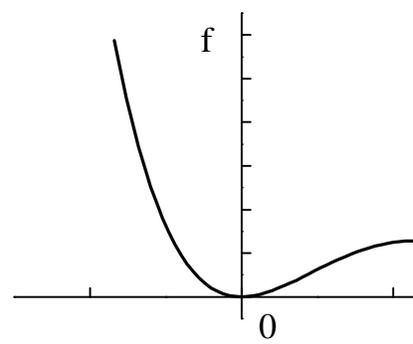
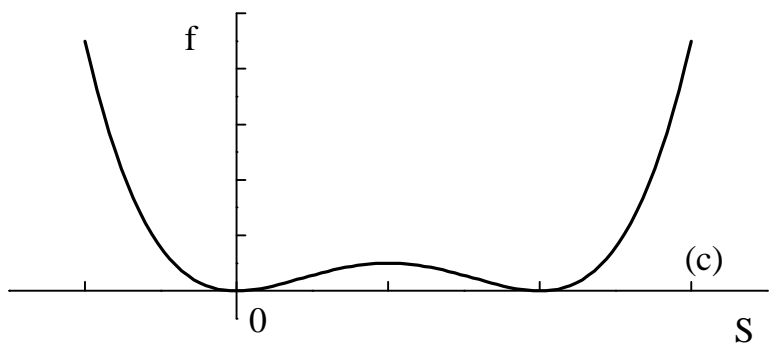
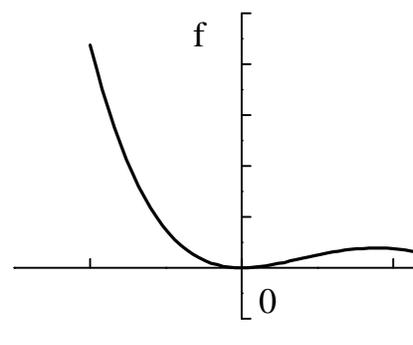
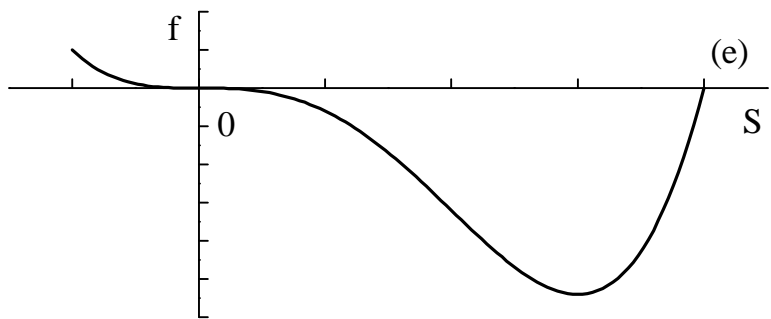
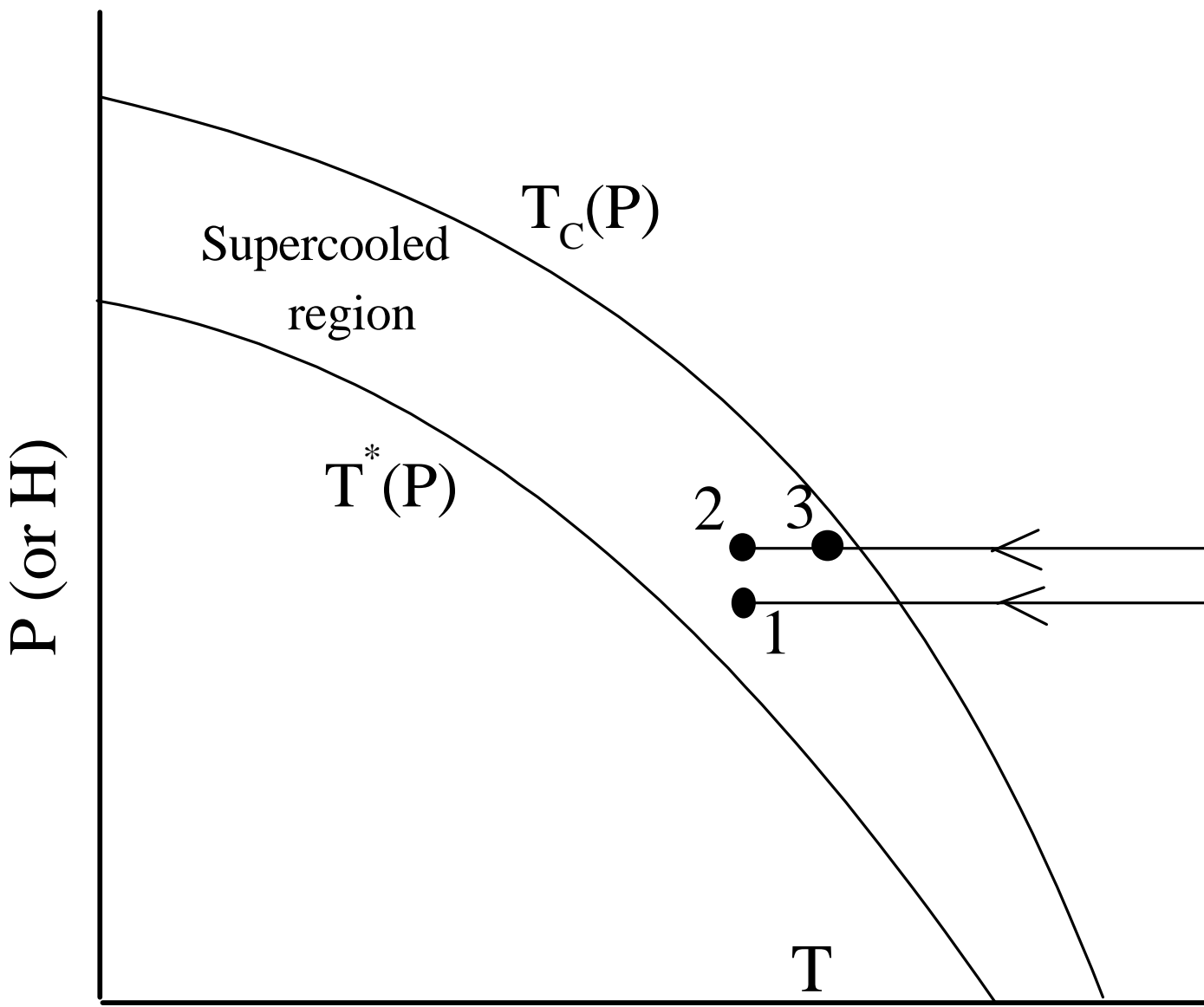
\end{document}